\documentstyle[12pt]{article}

\begin{document}

\begin{flushright}
IMSc/2001/07/39 \\
hep-th/0107255
\end{flushright}

\vspace{2ex}

\begin{center}

{\large \bf  }

\vspace{2ex}

{\large \bf  Some Consequences of the Generalised Uncertainty } \\

\vspace{2ex}

{\large \bf  Principle: Statistical Mechanical, Cosmological, } \\

\vspace{2ex}

{\large \bf  and Varying Speed of Light } \\

\vspace{8ex}

{\large  S. Kalyana Rama}

\vspace{3ex}

Institute of Mathematical Sciences, C. I. T. Campus, 

Taramani, CHENNAI 600 113, India. 
 
\vspace{1ex}

email: krama@imsc.ernet.in \\ 

\end{center}

\vspace{6ex}

\centerline{ABSTRACT}
\begin{quote} 
We study the dynamical consequences of Maggiore's unique
generalised uncertainty principle (GUP). We find that it leads
naturally, and generically, to novel consequences. In the high
temperature limit, there is a drastic reduction in the degrees
of freedom, of the type found, for example, in strings far above
the Hagedorn temperature. In view of this, the present GUP may
perhaps be taken as the new version of the Heisenberg
uncertainty principle, conjectured by Atick and Witten to be
responsible for such reduction. Also, the present GUP leads
naturally to varying speed of light and modified dispersion
relations. They are likely to have novel implications for
cosmology and black hole physics, a few of which we discuss
qualitatively.
\end{quote}

\vspace{2ex}



\newpage

{\bf 1.}  
Based on gedanken experiments in string theory
\cite{string} and black holes \cite{m1}, the
Heisenberg uncertainty principle is found to be modified to 
\begin{equation}\label{string}
\Delta x \ge \frac{\hbar}{\Delta p} 
+ (const) \; \frac{\lambda^2 \Delta p}{\hbar}
\end{equation} 
where $\Delta x$ and $\Delta p$ denote position and momentum
uncertainties, and $\lambda$ is a length parameter, given by
string length and Planck length in the above contexts. Under
certain assumptions, Maggiore has derived in \cite{m2,m3} 
unique generalised commutation relations (GCRs) which
lead to a generalised uncertainty principle (GUP) which, in
turn, leads to (\ref{string}) in a suitable limit. Under a
different set of assumptions it is possible to obtain more
general commutation relations, {\em e.g.} as in \cite{k},
which also lead to (\ref{string}). However, they are not unique.
Hence, in the following, we will consider Maggiore's
generalisation only, although our analysis is applicable to
other cases also.

The GCRs are kinematical. The dynamics is determined by
specifying a Hamiltonian $H$. To illustrate explicitly the
dynamical consequences of the GUP, we choose two candidates for
$H$ and study the statistical mechanics and the particle
dynamics of free particle systems obeying the GUP \footnote{ In
the case of the GCRs given by \cite{m2,m3}, some aspects of the
particle dynamics have been studied in \cite{m2}. In the case of
the GCRs given by \cite{k}, some aspects of the particle
dynamics, statistical mechanics, and black hole physics have
been studied in \cite{kmm}, \cite{lubo}, and \cite{lubobh}
respectively. See also \cite{ahluwalia}. }. We find that the
GUP leads naturally to many novel consequences.

In the high temperature limit, we find that there is a drastic
reduction in the degrees of freedom (d.o.f) and the
corresponding free energy is analogous to that found in certain
topological field theories and in strings far above the Hagedorn
temperature \cite{witten,witten2}. In this context, Atick and
Witten had indeed conjectured in \cite{witten} that a new
version of the Heisenberg uncertainty principle may be
responsible for such a reduction in the d.o.f. Here, we see that
such a reduction emerges naturally as a consequence of the GUP.
In view of this, the present GUP may perhaps be taken as the
conjectured new version of the Heisenberg uncertainty principle.

Another consequence of the GUP is the natural emergence of the
varying speed of light (VSL) and the modified dispersion
relations which, in turn, have non trivial implications for
cosmology and black hole physics \footnote{VSL theories were
first postulated in \cite{vsl1} and studied further in
\cite{vsl2,vslcov}. Their implications have been studied in
\cite{vsl2,vslcov,vsl3,vslbh}. The implications of modfied
dispersion relations for cosmology have been studied in
\cite{bb,niemeyer,mersini}.}. For one of the $H$'s considered
here, the VSL and the free energy together is likely to solve
the horizon problem in cosmology, as in \cite{magu}. The
corresponding VSL is likely to have novel implications for black
hole physics also.

The plan of the paper is as follows. In section 2, we present
the details of Maggiore's GCR and GUP, and the two candidate
Hamiltonians. In section 3, we study the statistical mechanics
and discuss the consequences. In section 4, we study the
particle dynamics and discuss the consequences. In section 5, we
present a brief summary and close by mentioning a few issues for
further study. 

{\bf 2.}  
The uncertainty principle (\ref{string}) can be thought of as
arising from a generalisation of the commutation relations
between the position operators $X_i$ and the momentum operators
$P_j$ in $d$-dimensional space where $i, j = 1, 2, \cdots, d$.  
Seeking the most general deformed Heisenberg algebra in 
$d = 3$, Maggiore has derived in \cite{m2,m3} 
the generalised commutation
relations (GCRs) between $X_i$ and $P_j$, determined uniquely
under the following assumptions: (i) The spatial rotation group
and, hence, the commutators ${[} J_i, J_j {]}$, ${[} J_i, X_j
{]}$, and ${[} J_i, P_j {]}$ are undeformed. (ii) The
translation group and, hence, the commutators ${[} P_i, P_j {]}$
are undeformed. (iii) The commutators ${[} X_i, X_j {]}$ and
${[} X_i, P_j {]}$ depend on a deformation parameter $\lambda$,
with dimension of length, and reduce to the undeformed ones in
the limit $\lambda \to 0$. The GCRs that follow uniquely from 
these assumptions are given by 
\begin{eqnarray}
{[} X_i, X_j {]} & = & - i \epsilon \hbar^2 
\lambda^2 \epsilon_{i j k} J_k 
\; \; , \; \; \;  
J_k = - i \epsilon_{k l m} P_l 
\frac{\partial}{\partial P_m}  \label{gxx} \\  
{[} X_i, P_j {]} & = & i \hbar \delta_{i j} f    
\; \; , \; \; \;  \; \; \; 
f = \sqrt{1 + \frac{\epsilon \lambda^2}{\hbar^2} 
(P^2 + m^2 c^2)} \label{gxp} 
\end{eqnarray} 
where $\epsilon = \pm 1$, $P^2 = \sum_i P_i^2$, and $\lambda$ is
a length parameter. The GCR (\ref{gxp}) then leads to the
generalised uncertainty principle (GUP) 
\begin{equation}\label{gup}
\Delta x_i \; \Delta p_j \ge \frac{\hbar}{2} \; 
\delta_{i j} \; \langle f \rangle \; . 
\end{equation} 
In the limit $\lambda^2 (p^2 + m^2 c^2) \ll \hbar^2$ and 
$\lambda \Delta p \stackrel{<}{_\sim} \hbar$, where $p^2$ 
is the eigenvalue of $P^2$, equation (\ref{gup}) reduces to
(\ref{string}). See \cite{m2,m3} for details. 

In the following, we consider $d$-dimensional space \footnote{
With the right hand side of (\ref{gxx}) written explicitly in
terms of $P_i$ and $\frac{\partial}{\partial P_j}$, equations
(\ref{gxx}) - (\ref{gup}) are valid in $d$-dimensional space
also, as can be easily verified.}. We set $\hbar = c = 1$ 
unless indicated otherwise. The case $\epsilon = - 1$ implies 
a bound $\lambda^2 (p^2 + m^2) < 1$, whose physical
significance is not clear. Hence, we will consider the case
$\epsilon = 1$ only.  Also, we consider non rotating systems
only and, therefore, set $J_k = 0$ in (\ref{gxx}). The only
nontrivial GCR is then given by (\ref{gxp}) with 
\begin{equation}\label{f} 
f = \sqrt{1 + \lambda^2 (P^2 + m^2)} 
\; , \; \; \; \; P^2 = \sum_{i = 1}^d P_i^2 \; . 
\end{equation}

We study the consequences of the GCR (\ref{gxp}), equivalently
of the GUP (\ref{gup}), with $f$ given by (\ref{f}). It is
important to note that the commutation relations are kinematical
only. The dynamics is determined by the Hamiltonian $H$ which,
therefore, needs to be specified.  In the following, we assume
that $H$ depends on $P$ only through the rotationally invariant
combination $\sqrt{P^2 + m^2}$. Furthermore, for the sake of
simplicity, we consider here only the free particle case, for
which $H$ is independent of $X$.

To illustrate the non trivial consequences of the GUP
(\ref{gup}), we consider two choices for $H$. They are given by 
\begin{eqnarray}
H' = 1 & \; \; \; \longleftrightarrow \; \; \; & 
H = \sqrt{P^2 + m^2} \label{h1} \\ 
f H' = 1 & \; \; \; \longleftrightarrow \; \; \; & 
Sinh \; \lambda H  = 
\lambda \sqrt{P^2 + m^2}  \label{h2}
\end{eqnarray} 
where $H'$ is the derivative of $H$ with respect to 
$\sqrt{P^2 + m^2}$ and $f$ is given by (\ref{f}). The
Hamiltonian $H$ in (\ref{h1}) is, perhaps, the simplest and a
natural choice; $H$ in (\ref{h2}) is obtained in \cite{m2} from
the first Casimir operator and, hence, is also a natural choice
from group theoretic point of view.

{\bf 3.}   
Consider the statistical mechanics of a system of free particles
confined in a $d$-dimensional volume $V$, which obey the GUP
(\ref{gup}) with $f$ given by (\ref{f}). The calculations in the
microcanonical or canonical ensemble approach are complicated.
But they are simple in the grand canonical ensemble approach
which we, therefore, use.

In the standard case where $\lambda = 0$, the one-particle phase
space measure is given by $h^{- d}$ where $h$ is the Planck's
constant. Physically, this is because the phase space is divided
into cells of volume $h^d$ as a consequence of the Heisenberg
uncertainty principle. Various phase space integrals will be of
the form $\int d^d x d^d p \; h^{- d} \; (*)$ where $x$ and $p$
denote the eigenvalues of $X$ and $P$. In the present case, the
particles are assumed to obey the GUP (\ref{gup}). Consequently,
the phase space must be divided into cells of volume $h^d
f^d$. Then, the one-particle phase space measure is given by
$h^{- d} f^{- d}$, and various phase space integrals are of the
form $\int d^d x d^d p \; h^{- d} f^{- d} \; (*)$.

The function $f$ and, in the free particle case, the Hamiltonian
$H$ depend on $P$ only. $x$-integration then simply gives a
volume factor $V$. Writing $p$ in terms of energy $E$, which is
the eigenvalue of $H$, the phase space integrals can be written
as 
\begin{equation}\label{measuredefn}
\int \frac{d^d x d^d p}{h^d f^d} \; (*) 
\equiv \int d E \; g(E) \; (*) \; . 
\end{equation} 
The measure $g(E)$ is the analog of the one-particle density of
states. It can be easily calculated and is given by 
\begin{equation}\label{gh} 
g(E) = \frac{{\cal C} \; p^{d - 2}}{f^d \; E'} \; 
\sqrt{p^2 + m^2} \; \; , \; \; \; {\rm where} \; \; \; 
{\cal C} \equiv \frac{\Omega_{d - 1} V}{h^d} \; , 
\end{equation}  
$\Omega_{d - 1}$ is the area of a unit $(d - 1)$-dimensional
sphere, and $E'$ is the derivative of $E$ with respect to 
$\sqrt{p^2 + m^2}$. In equation (\ref{gh}), $p$ is to be 
expressed in terms of $E$.

Consider the grand canonical ensemble. It can be easily verified
that the definitions of, and the relations between, various
thermodynamical quantities all remain unchanged, with $g(E)$
given as in (\ref{gh}) \cite{greiner}. Thus, we have 
\begin{equation}\label{thermod}
- \beta F = \beta {\cal P} V = ln {\cal Z} = 
\frac{1}{a} \int_0^\infty d E g(E) \; 
ln (1 + a e^{- \beta (E - \mu)})  
\end{equation}   
where we have used the standard notation: $\beta = T^{- 1}$ is
the inverse temperature, $F$ is the free energy, ${\cal P}$ is
the pressure, ${\cal Z}$ is the grand canonical partition
function, and $\mu$ is the chemical potential. Also, 
$a = - 1, 0$, or $+ 1$ depending on whether the particles obey,
respectively, Bose-Einstein, Maxwell-Boltzmann, or Fermi-Dirac
statistics.  When $a = 0$, $ln {\cal Z}$ is to be evaluated in
the limit $a \to 0$. Various thermodynamical quantities can then
be calculated using (\ref{thermod}): for example, the internal
energy $U = - \frac{\partial ln {\cal Z}}{\partial \beta}$, the
particle number $N = \frac{\partial ln {\cal Z}}{\beta \partial
\mu}$, the entropy $S = \beta \; (U + {\cal P} V - \mu N)$, etc.

It is clear from the above formulae, or from physical
arguments, that the effect of $\lambda$ will be considerable
only when the temperature/energy is of ${\cal O}(\lambda^{- 1})$ 
or higher. Therefore, the limit of interest here is the high
temperature limit $\beta \ll \lambda$. Also, we expect that
$\lambda$ is extremely small, in particular, $\lambda m \ll
1$. (For example, $\lambda \simeq string/Planck$ $length$.)
Furthermore, in the high temperature limit, the statistics is
irrelevent. Therefore, for the sake of simplicity, we set 
$m = 0$ and $a = 0$ in the following.  Then $\mu = 0$ since 
$m = 0$. One then obtains $U, S, N$ etc. in terms of $V$ and $T$
using (\ref{thermod}) \cite{greiner}.

To proceed further, $p$ and $f$ in equation (\ref{gh}) are to be
expressed in terms of $E$, for which an explicit form of $H$ is
required. We consider $H$ given in (\ref{h1}) and (\ref{h2}),
with $m = 0$. A simple algebra then shows that $g(E)$ is given
by 
\begin{eqnarray} 
H' = 1 & \; \; \; \longleftrightarrow \; \; \; & 
g(E) = \frac{{\cal C} \; E^{d - 1}} 
{(1 + \lambda^2 E^2)^{\frac{d}{2}}} \label{gh1} \\ 
f H' = 1 & \; \; \; \longleftrightarrow \; \; \; & 
g(E) = \frac{{\cal C}}{\lambda^{d - 1}} \; 
(tanh \; \lambda E)^{d - 1}  \; . \label{gh2} 
\end{eqnarray} 
The partition function ${\cal Z}$ and other quantities can now
be evaluated in closed form in terms of special functions, as
described in the Appendix. 

We expect to obtain the standard results in the limit $\beta \gg
\lambda$ and to obtain the non trivial features, if any, in the
limit $\beta \ll \lambda$\footnote{In both of these limits, it
is easier to evaluate the integral in (\ref{thermod}) directly,
using equations (\ref{gh1}) and (\ref{gh2}) and appropriate
Taylor expansions, than to work with the special functions.}. A
simple calculation in the limit $\beta \gg \lambda$ shows that,
to the leading order in $\frac{\lambda}{\beta}$, the
thermodynamical quantities are independent of $\lambda$, and are
indeed the standard ones for a gas of massless free particles in
$d$-dimensional space obeying the Heisenberg uncertainty
principle. They are given by \cite{greiner} 
\begin{equation}\label{ans0} 
- \beta F = \frac{{\cal C} (d - 1)!}{\beta^d} 
\; \; , \; \; \; 
U = \frac{{\cal C} d !}{\beta^{d + 1}} 
\; \; , \; \; \; 
S = \frac{{\cal C} (d + 1)!}{d \beta^d} \; . 
\end{equation}
Note that the free energy has the behaviour 
$|\beta F| \sim T^d$.

Consider the limit $\beta \ll \lambda$. The results now depend
on whether $g(E)$ is given by equation (\ref{gh1}) or
(\ref{gh2}).  In the case where $g(E)$ is given by (\ref{gh1}),
the thermodynamical quantities are given, to the leading order
in $\frac{\beta}{\lambda}$, by 
\begin{equation}\label{ans1} 
- \beta F = const + 
\frac{{\cal C}}{\lambda^d} \; ln (\lambda T) 
\; \; , \; \; \; 
U = \frac{{\cal C}}{\beta \lambda^d} 
\; \; , \; \; \; 
S =  const + 
\frac{{\cal C}}{\lambda^d} \; (1 + ln (\lambda T))  \; . 
\end{equation}

Note that in the high temperature limit $\beta \ll \lambda$, the
free energy has the behaviour $|\beta F| \sim (const) + \ln T$
for any value of $d$. This indicates a drastic reduction in the
degrees of freedom (d.o.f). Such reduction may be possible, in
the context of Heisenberg uncertainty principle, if a continuum
field theory is replaced by a lattice theory, with a finite
number of bose oscillators at each site \cite{witten}; or, in
certain topological theories \cite{witten2} with general
covariance restored at short distances \cite{witten}. Here, we
see that such a behaviour emerges naturally as a consequence of
the GUP (\ref{gup}), for systems whose Hamiltonian $H$ is given
by (\ref{h1}).

The thermodynamical relations (\ref{ans1}), which, upto
polarisation factors, are valid for photons also, may have
interesting cosmological consequences. In a recent paper
\cite{magu}, the authors analyse the thermodynamical behaviour
of photons in the framework of non commutative geometry,
postulating a model dependent varying speed of light (VSL).
Amazingly, although their set up bears no discernible relation
to the present one, the equation of state 
\begin{equation}\label{eos}
U = \frac{{\cal C}}{\beta \lambda^d} 
\; \; , \; \; \; 
\frac{{\cal P}}{U} = (const) + ln (\lambda T) 
\end{equation} 
in the present case, obtained from (\ref{thermod}) and
(\ref{ans1}), is essentially the same as that obtained in
\cite{magu} (equations (38) and (39) of \cite{magu}). It is
shown in \cite{magu} that such an equation of state, with an
additional ingredient to be mentioned in the next section,
solves the horizon problem.

Consider the limit $\beta \ll \lambda$, now in the case where
$g(E)$ is given by (\ref{gh2}). The thermodynamical quantities
are given, to the leading order in $\frac{\beta}{\lambda}$, by 
\begin{equation}\label{ans2} 
- \beta F = const + 
\frac{{\cal C}}{\lambda^{d - 1} \beta} 
\; \; , \; \; \; 
U = \frac{{\cal C}}{\lambda^{d - 1} \beta^2} 
\; \; , \; \; \; 
S = const + 
\frac{2 \; {\cal C}}{\lambda^{d - 1} \beta} \; . 
\end{equation}

Note that in the high temperature limit $\beta \ll \lambda$, the
free energy has the behaviour $|\beta F| \sim (const) + T$ for
any value of $d$. This indicates a drastic reduction in the
d.o.f. Precisely such a free energy, and hence such a reduction,
has been found in \cite{witten} in the case of the strings at
temperatures far above the Hagedorn temperature. In this
context, Atick and Witten had indeed conjectured in
\cite{witten} that a new version of the Heisenberg uncertainty
principle may be responsible for such a reduction in the
d.o.f. Here, we see that such a drastic reduction in the d.o.f
emerges naturally as a consequence of the GUP (\ref{gup}), for
systems whose Hamiltonian $H$ is given by (\ref{h2}). In view of
this, the GUP (\ref{gup}) may perhaps be taken as the
conjectured new version of the Heisenberg uncertainty principle.

The physical origin of the reduction in the d.o.f is easy to
understand. When the system obeys the GUP (\ref{gup}), the
volume of the phase space cells is $h^d f^d$. It grows at high
temperatures/energies and, since the function $f$ is given by
(\ref{f}), $\sim p^d$. Consequently, the number of available
cells is enormously reduced, compared with the standard
case. This, essentially, is the origin of the reduction in the
d.o.f seen above. The precise amount of reduction depends on the
choice of the Hamiltonian $H$.  For the $H$ given by $f^\alpha
H' = 1$, where $\alpha = 0, 1$,\footnote{For $\alpha > 1$, $H$
becomes bounded, {\em i.e.} its eigenvalue $E \to (const)$ as 
$p \to \infty$, a behaviour whose significance is not clear.} 
it can be seen from equation (\ref{gh}) that the reduction is
such that the resulting d.o.f are equal to that of an
$\alpha$-dimensional system obeying the Heisenberg uncertainty
principle and with an effective volume (effective number of
sites in the $\alpha = 0$ case) 
$V_\alpha \sim V \lambda^{\alpha - d}$. This is precisely the
result seen explicitly in equation (\ref{ans1}) for $\alpha = 0$
and in equation (\ref{ans2}) for $\alpha = 1$.

{\bf 4.} 
Consider the dynamics of particles which obey the GUP
(\ref{gup}), equivalently the GCR (\ref{gxp}), with $f$ given 
by (\ref{f}). We first note in passing that the time energy
uncertainty relation remains unchanged. The derivation proceeds
in the standard way, {\em e.g.} as in \cite{messiah}, with the
result that 
\begin{equation}\label{te}
\Delta t_Q \; \Delta E \ge \frac{\hbar}{2}
\end{equation}
where $\Delta t_Q \simeq \Delta Q \left(\frac{d Q}{d t}
\right)^{- 1}$ is the time uncertainty, time being measured by
measuring the variation of an observable $Q$, which has no
explicit time dependence and, hence, obeys $\frac{d Q}{d t} =
\frac{i}{\hbar} \; {[} H, Q {]}$ where $H$ is the Hamiltonian.

We now define the velocity operator $V_i$ by (see \cite{m2}
also) 
\begin{equation}\label{vi}
V_i \equiv \frac{d X_i}{d t} = 
\frac{i}{\hbar} \; {[} H, X_i {]} \; . 
\end{equation} 
In the case of a non rotating system for which ${[} X_i, X_j
{]} = 0$, or a free particle system for which $H$ is
independent of $X_i$, one obtains using (\ref{gxp}) that
\begin{equation}\label{vel}
V_i = \frac{f H' P_i}{\sqrt{P^2 + m^2}}  
\end{equation} 
where $H'$ is the derivative of $H$ with respect to 
$\sqrt{P^2 + m^2}$. Denoting the eigenvalue of $V_i$ by 
$v_i$, the speed $v$ of a particle with mass $m$ is given by 
\begin{equation}\label{v}
v = \left( \sum_{i = 1}^d v_i^2 
\right)^{\frac{1}{2}} = 
\frac{p f E'}{\sqrt{p^2 + m^2}} \; . 
\end{equation}

In Quantum Mechanics, with $\hbar = 1$, the energy eigenfunction
will have a time dependence $e^{- i \omega t}$ where $\omega =
E$, the eigenvalue of $H$. Its group velocity $\frac{d \omega}{d
k}$ can then be identified naturally with the velocity $v$ in
(\ref{v}) where, now, $E = \omega$. This leads to a modified
dispersion relation, given by 
\begin{equation}\label{disp}
\frac{d \omega}{d k} = \frac{p f E'}{\sqrt{p^2 + m^2}}
\end{equation} 
where one sets $E = \omega$ on the right hand side. Note that 
equation (\ref{disp}) can be thought of as defining the
wave number $k$ in the position space. Indeed, in $d = 1$, $k$
is the eigenvalue of the operator $K$ defined by $f \; \frac{d
K}{d P} = 1$, so that ${[} X, K {]} = i$. Equation (\ref{disp})
then follows since $\frac{d H}{d K} = \frac{P H'}{\sqrt{P^2 +
m^2}} \; \frac{d P}{d K} = \frac{P f H'}{\sqrt{P^2 + m^2}}$.

Furthermore, the speed of light, denoted by $C$, can be
identified naturally with the speed of a particle with mass 
$m = 0$. Equation (\ref{v}) then gives 
\begin{equation}\label{c}
C = f E' 
\; \; \; \; {\rm and} \; \; \; \; 
v \le C  
\; . 
\end{equation} 

In equations (\ref{v}), (\ref{disp}), and (\ref{c}) $p$ and $f$
are to be expressed in terms of $E$, for which an explicit form
of $H$ is required. For $H$ given by (\ref{h1}) we have,
after a simple integration, 
\begin{eqnarray}
v & = & \frac{\sqrt{(E^2 - m^2) (1 + \lambda^2 E^2)}}{E}
\; \; , \; \; \; \; 
C = \sqrt{1 + \lambda^2 E^2} \label{vc1} \\
\lambda^2 \omega^2 & = & (1 + \lambda^2 m^2) \; \; 
Sinh^2 \; \lambda k + \lambda^2 m ^2 \; . \label{disp1} 
\end{eqnarray} 
For $H$ given by (\ref{h2}) (see \cite{m2} also) 
we have, after a simple integration,  
\begin{eqnarray} 
v & = & \frac{\sqrt{Sinh^2 \; \lambda E - \lambda^2 m ^2}} 
{Sinh \; \lambda E} 
\; \; , \; \; \; \; 
C = 1 \label{vc2} \\
Cosh \; \lambda \omega & = & \sqrt{1 + \lambda^2 m^2} \; \; 
Cosh \; \lambda k \; . \label{disp2} 
\end{eqnarray} 

We now discuss the physical significance of these results.
Consider first the modified dispersion relation (\ref{disp}). 
For $\lambda = 0$, it reduces to $\omega^2 = k^2 + m^2$. For
$\lambda \ne 0$, the modification is generically non trivial.
The exception is when the Hamiltonian $H$ is given by
(\ref{h2}), in which case the modification is only
marginal. These can be seen explicitly by considering the high
energy `transplanckian' limit $\lambda \omega \gg 1$ of
equations (\ref{disp1}) and (\ref{disp2}).

Recent studies \cite{bb,niemeyer} have suggested that such
`transplanckian' modifications of dispersion relation may have
observable consequences for the density perturbations that arise
during inflation. In these studies, the modified dispersion
relations need to be postulated. Here, however, we see that
generically the GUP (\ref{gup}) leads naturally to modified
dispersion relations. It is clearly of interest to study their
consequences for the density fluctuations that arise during
inflation.

Consider the speed of light $C$. It is clear from equation
(\ref{c}) that, generically, $C$ is varying and is a non trivial
function of energy $E$. The exception is when the Hamiltonian
$H$ is given by (\ref{h2}), in which case $C = 1$. Such `varying
speed of light' (VSL) theories have been extensively studied
\cite{vsl1,vsl2,vslcov}, and found to have non trivial
implications for cosmology \cite{vsl3} and black hole physics
\cite{vslbh}. In these theories, VSL needs to be postulated.
Here, however, we see that generically the GUP (\ref{gup}) leads
naturally to VSL. It is clearly of interest to study its
consequences, which are likely to be non trivial.

For example, $C$ given by (\ref{vc1}) increases with energy
$E$. This is precisely the ingredient, alluded to below equation
(\ref{eos}), that is necessary, but postulated, in \cite{magu}
to solve the horizon problem.  Here, it arises naturally.
Moreover, a preliminary analysis shows that the VSL, given by
(\ref{vc1}), and the photon distribution at energies $E \gg
\lambda^{- 1}$, deriveable from (\ref{thermod}) and (\ref{gh1}),
both have the right behaviour needed to solve the horizon
problem, as in \cite{magu}, but within the present framework. 
A detailed analysis, however, is beyond the scope of the present
letter.

Equations (\ref{vc1}) are likely to have novel implications for
black hole physics also. The horizon of a black hole can naively
be thought of as the place where the escape velocity $= 1$, in
units where $c = 1$. Particles can then escape from, or from
even inside, the horizon if their energy $E$ is sufficiently
high since their speed can then be $> 1$.  This may, therefore,
provide a mechanism for the transfer of information from inside
the horizon to the outside, a process for which no mechanism is
known at present \cite{page}.

Perhaps more correctly, the horizon is to be thought of as the
place from where nothing can escape. Then the escape velocity at
the horizon must be infinite since $v$ and $C \to \infty$ as $E
\to \infty$. Very likely, therefore, the horizon size must be
infinitesimally small or, perhaps equivalently, no black holes
can form \footnote{In this context, note that a certain class of
VSL theories are shown in \cite{vslcov} to be equivalent to
generalised Brans-Dicke theories. For a certain class of the
later theories, it is argued in \cite{kal} that black holes are
unlikely to form. For a discussion of black hole physics in VSL
theories from another point of view, see \cite{vslbh}.}. These
implications for the black hole physics of the GUP (\ref{gup}),
with $f$ given by (\ref{f}) and the Hamiltonian $H$ given by
(\ref{h1}), are very interesting but the present arguments are,
admittedly, qualitative. Unfortunately, in the absence of a
Lorentz and/or general coordinate invariant formulation of the
GUP, these issues cannot be addressed rigorously. 

{\bf 5.}  
In summary, we have studied the physical consequences of the GUP
(\ref{gup}) that follows from the GCR (\ref{gxp}) which is
determined uniquely by Maggiore under a set of assumptions. We
studied the statistical mechanics and the particle dynamics of
systems obeying the GUP (\ref{gup}) and found novel consequences
arising in a natural way. For example, the GUP leads naturally
to free energies of the form found in certain topological field
theories and in strings far above the Hagedorn temperature. It
also leads naturally to VSL and to modified dispersion
relations. Among other things, these features are likely to
solve the horizon problem in cosmology, and may provide novel
insights into black hole physics also.

There are numerous issues that require further study. We close
by mentioning a few of them. (i) Understanding the physical
significance of the bound $\lambda^2 (p^2 + m^2) < 1$ in the
$\epsilon = - 1$ case. (ii) Finding the physical principle, if
any, which selects a given Hamiltonian $H$, {\em e.g.} the one
given by $f^\alpha H' = 1$ for a given value of $\alpha$.  
(iii) Finding a Lorentz and/or general coordinate invariant
formulation of the GUP which is crucial, for example, for the
study of black hole physics. (iv) Exploring relations, if any,
between GUP and string theory, topological field theory, and VSL
theories. That such a relation may exist is, perhaps, indicated
by equations (\ref{ans2}), (\ref{ans1}), and (\ref{c}).


\vspace{2ex}

\centerline{{\bf Appendix}}

\vspace{1ex}

The partition function in (\ref{thermod}), with $m = a = 0$, can
be obtained in a closed form in terms of special functions both
for the cases where $H' = 1$, and where $f H' = 1$. We need to
evaluate integrals of the form 
\begin{eqnarray} 
I_{m, n} & = & \int_0^\infty d t \; 
t^n (\alpha^2 + t^2)^{-\frac{m}{2}} \; e^{- s t} 
\; \; \; {\rm for} \; \; H' = 1 \label{imn} \\
J_{m, n} & = & \int_0^\infty d t \; 
t^n (tanh \; t)^m \; e^{- s t}  \; \; \; \; \; \; 
{\rm for} \; \; f H' = 1 \; , \label{jmn} 
\end{eqnarray}
where $t = \lambda E$, $s = \frac{\beta}{\lambda}$, and
$\alpha^2 = 1$. It is easy to see that 
\begin{eqnarray*}
I_{2 k, n} & = & \frac{(- 1)^{n + k - 1}}{(k - 1)!} \;
\left( \frac{d}{d s} \right)^n \; 
\left( \frac{d}{d \alpha^2} \right)^{k - 1} I_{2, 0} \\ 
I_{2 k + 1, n} & = & 
\frac{(- 1)^{n + k} \; 2^{k + 1}}{(2 k - 1)!!} \;
\left( \frac{d}{d s} \right)^n \; 
\left( \frac{d}{d \alpha^2} \right)^{k + 1} I_{- 1, 0} \\
J_{m, n} & = & (- 1)^n \left( \frac{d}{d s} \right)^n 
J_{m, 0} \; .
\end{eqnarray*} 
Clearly, 
$J_{0, 0} = \frac{1}{s}$. Moreover, 
\begin{eqnarray*}
I_{2, 0} & = & \frac{1}{\alpha} \; 
( ci(\alpha s) sin(\alpha s) 
- si(\alpha s) cos(\alpha s) ) \\
I_{- 1, 0} & = & \frac{\pi \alpha}{2 s} \; 
( {\bf H}_1(\alpha s) - {\bf Y}_1(\alpha s) ) \\
J_{1, 0} & = & \beta(\frac{s}{2}) - \frac{1}{s} 
\end{eqnarray*} 
where $I_{2, 0}$ is given in equation (3.354.1) of \cite{gr};
$J_{1, 0}$ in (3.541.7) of \cite{gr}; and $I_{- 1, 0}$ is
obtained using equation (4.2.27) of \cite{erdelyi} and the
properties of the Laplace transform of the derivative of a
function. Here, $ci$ and $si$ are the cosine and sine integrals
respectively (see section (8.23) of \cite{gr}), ${\bf H}_1$ is
the Struve's function, ${\bf Y}_1$ is the Bessel function of the
second kind, and the function $\beta(x)$ is related to the
derivatives of the gamma function (see section 8.37 of 
\cite{gr}).

We now obtain a recursion relation for $J_{m, 0}$. We have, for
$m \ne 1$, 
\begin{equation}\label{tm}
T_{m}(x) \equiv \int_0^x d x' (tanh \; x')^m = 
- \frac{(tanh \; x)^{m - 1}}{m - 1} + T_{m - 2}(x) \; . 
\end{equation}
Thus, after a partial integration in $J_{m, 0}$ and using 
$T_m(0) = 0$, one obtains 
\begin{equation}\label{jm0}
J_{m, 0} = s \int_0^\infty d x T_m(x) e^{- s x} \; . 
\end{equation}
Equations (\ref{jmn}), (\ref{tm}), and (\ref{jm0}) now lead to
the recursion relation 
\begin{equation}
J_{m, 0} = - \frac{s}{m - 1} \; J_{m - 1, 0} 
+ J_{m - 2, 0} \; , 
\end{equation}  
using which $J_{m, 0}$ can be obtained for $m \ge 2$. Using
these formulae, the partition function and other quantities can
be evaluated in closed form.


\vspace{2ex}



\end{document}